# Current-Controlled Topological Magnetic Transformations in a Nanostructured Kagome Magnet

*Wensen Wei[#], Jin Tang[#*], Yaodong Wu, Yihao Wang, Jialiang Jiang, Junbo Li, Y. Soh, Yimin Xiong, Mingliang Tian, and Haifeng Du[*]*

Dr. W. Wei, Dr. J. Tang, Y. Wu, Dr. Y. Wang, J. Jiang, J. Li, Prof. Y. Xiong, Prof. M. Tian, Prof. H. Du
Anhui Province Key Laboratory of Condensed Matter Physics at Extreme Conditions, High Magnetic Field Laboratory, HFIPS, Anhui, Chinese Academy of Sciences, Hefei, 230031, China
E-mail: tangjin@hmfl.ac.cn; duhf@hmfl.ac.cn
[#]W.W. and J.T. contributed equally to this work.

Y. Wu
Key Laboratory for Photoelectric Detection Science and Technology of Education Department of Anhui Province, and School of Physics and Materials Engineering, Hefei Normal University, Hefei, 230601, China

Prof. Y. Soh
Paul Scherrer Institute, 5232, Villigen, Switzerland

Prof. M. Tian
School of Physics and Materials Science, Anhui University, Hefei, 230601, China

Prof. H. Du
Institutes of Physical Science and Information Technology, Anhui University, Hefei, 230601, China

**Abstract**

Topological magnetic charge $Q$ is a fundamental parameter that describes the magnetic domains and determines their intriguing electromagnetic properties. The ability to switch $Q$ in a controlled way by electrical methods allows for flexible manipulation of electromagnetic behavior in future spintronic devices. Here we report the room-temperature current-controlled topological magnetic transformations between $Q = -1$ skyrmions and $Q = 0$ stripes or type-II bubbles in a kagome crystal $Fe_3Sn_2$. We show that the reproducible and reversible skyrmion-bubble and skyrmion-stripe transformations can be achieved by tuning the density of nanosecond pulsed current of the order of $\sim 10^{10}$ A m$^{-2}$. Further numerical simulations suggest that spin-transfer torque combined with Joule thermal heating effects determine the current-induced topological magnetic transformations.

**Introduction**

Magnetic skyrmions, which are topologically non-trivial magnetic swirls that are potentially exploitable in future spintronic devices, have been actively researched in the last decade[1-8]. The key parameter describing magnetic skyrmions is the topological magnetic charge $Q$, which is defined as $Q = 1/(4\pi) \int \mathbf{m} \cdot (\frac{\partial \mathbf{m}}{\partial x} \times \frac{\partial \mathbf{m}}{\partial y}) \, dx dy$ and essentially counts how many times the magnetization vector field $\mathbf{m}$ winds around a unit sphere[1]. Emergent electromagnetic properties of skyrmions originated from topological charge $Q$, e.g., topological Hall effects and skyrmion Hall effects, have been revealed[3, 9-11]. Therefore, manipulation of topological magnetic charges in magnetic materials by electrical methods is of long-term interest for future skyrmion-electronic devices[12]. The switching of topological magnetic charge can be realized by transformations between skyrmions with an integer charge $Q$ and topologically trivial magnetic states with zero $Q$, such as ferromagnet[13-17], helix or stripes[18-25], and bubbles[26-33]. Previous studies have realized the current-induced skyrmion-ferromagnet transformations[34, 35]. However, thus far, current-induced skyrmion–stripe (or helix) and skyrmion–bubble transformations which occur along with the manipulation of topological charges in a reproducible and reversible way have not been experimentally demonstrated.

Here, we show room-temperature current-controlled manipulation of topological magnetic charge in an achiral kagome crystal $Fe_3Sn_2$ by skyrmion–bubble and skyrmion–stripe transformations. The uniaxial magnetic crystal $Fe_3Sn_2$ possesses two types of localized magnetic objects, *i.e.*, topologically nontrivial skyrmion bubbles with an integer charge $Q$ and topologically trivial bubbles with zero-$Q$[26-28]. Because skyrmion bubbles are topologically equivalent to chiral skyrmions[26-32], the two classes of magnetic textures share the same electromagnetic properties which are determined by topology $Q$. Thus, skyrmion bubbles are also called skyrmions in achiral uniaxial ferromagnets[30-33]. Here, for simplicity, skyrmion bubbles and topologically trivial bubbles in Fe$_3$Sn$_2$ are named skyrmions and bubbles, respectively. Similar multi-$Q$ magnetic objects are also observed in antiskyrmion-hosting Mn$_{1.4}$Pt$_{0.9}$Pd$_{0.1}$Sn and Fe$_{1.9}$Ni$_{0.9}$Pd$_{0.2}$P alloys[36, 37]. However, previous magnetic field-controlled topological magnetic transformations among these multi-$Q$ objects are not compatible for future electrically controlled devices[31, 32, 36, 37], therefore, current-controlled transformations remain elusive.

**Results and discussions**

The kagome crystal Fe$_3$Sn$_2$ is a room-temperature uniaxial ferromagnet with magnetization easy axis along the [001] axis[38, 39]. The formation of magnetic objects in Fe$_3$Sn$_2$ originates from the competition of uniaxial anisotropy, dipole–dipole interaction, and Zeeman energy[26, 27]. Skyrmions with closure domain wall magnetizations are typically stabilized at a normal field to minimize the dipole–dipole interaction. In contrast, under a tilted magnetic field, the domain wall magnetizations tend to align along the in-plane components of magnetic field to minimize Zeeman energy, resulting in the formation of bubbles. Skyrmions and bubbles in achiral uniaxial magnets can also form as closely-packed hexagonal lattices[31]. However, skyrmion lattice and bubble lattice exhibit in different styles. Two of the six sides of a unit cell of the hexagonal bubble lattice tend to align along the in-plane field orientation to minimize the dipole–dipole interaction energy (Supplemental Figure 1). In contrast, the six sides of hexagonal skyrmion lattice have no preferred orientations because of the closure domain wall magnetizations of skyrmions. Because of the absence of chiral Dzyaloshinskii–Moriya interaction (DMI) in the centrosymmetric Fe$_3$Sn$_2$ crystal, skyrmions with two types of helicity are energetically equivalent (**Figure 1**). Of note, skyrmions with two different helicities and same polarity at a given out-of-plane magnetic field $B$ share the same $Q$ values (Figure 1d). Because Lorentz-TEM is only sensitive to the projected in-plane magnetizations, and skyrmions with counterclockwise and clockwise rotations of in-plane magnetic components are shown by white and black dots under the under-focus conditions, respectively. Bubbles shown by arc-shaped Fresnel contrasts prefer to be stabilized in the presence of in-plane field with their cylinder domain wall magnetizations oriented towards the in-plane field. A previous study has shown the emergent dynamic response of skyrmions to current such as helicity reversal of skyrmions in $Fe_3Sn_2$[28]. However, random helicity reversals of skyrmions occur not in a controlled way. In addition, the topological magnetic charge $Q$ remains unchanged during helicity reversal, which suggests that the electromagnetic properties determined by $Q$ were not affected. Here, we first show room-temperature topological magnetic transformations between skyrmions with $Q = -1$ and bubbles with $Q = 0$ in $Fe_3Sn_2$(001) thin plates.

A ~150-nm-thick $Fe_3Sn_2$(001) microdevice was chosen for the *in-situ* Lorentz transmission electron microscopy (TEM) investigation of current-induced magnetic domain dynamics (Figure 1a)[40]. The pulse width and periodicity of the current are set to 120 ns and 1 s, respectively. In the presence of high magnetic field, we mainly obtain skyrmions in the $Fe_3Sn_2$ thin plate. Similar to chiral skyrmions, dense skyrmions in $Fe_3Sn_2$ arrange to form nearly hexagonal packed lattices (Figure 1c). We explore the current density dependence of magnetic dynamics in the $Fe_3Sn_2$ thin plate (Figure 1c, Supplemental Figure 2, and Supplemental video 1). At $B \sim 500$ mT with a tilted angle $\alpha \sim 2.0°$ with $\alpha$ defined as the angle between the out-of-plane [001] axis and external field, we achieve current-induced random helicity reversal of skyrmions at $j \sim 3.16 \times 10^{10}$ A m$^{-2}$ (Supplemental Figure 3), similar to those observed previously[28]. When continuously applying the pulsed current, we slowly increased the current density with a step time interval of ~60 s (~60 current pulses) per increment of current density. For $j < 3.88 \times 10^{10}$ A m$^{-2}$, we observed not only the helicity reversal of skyrmions but also a slow variation of skyrmion counts and bubble counts (Figure 1b). In the current density range of ~3.88–4.16 × 10$^{10}$ A m$^{-2}$, a rapid topological magnetic transition from mainly skyrmions to bubbles occurs (Figures 1b, c and Supplemental video 1). At $j > \sim 4.16 \times 10^{10}$ A m$^{-2}$, we observed mainly the rearrangement of the bubble lattice. Interestingly, a well-reversed topological magnetic transition from the bubble lattice to skyrmion lattice is achieved by decreasing the current density from ~4.78 to $3.16 \times 10^{10}$ A m$^{-2}$ (Figure 1b, c and Supplemental video 1). The skyrmion count decreased from ~150 ($Q = -150$) at $j \sim 3.16 \times 10^{10}$ A m$^{-2}$ to ~ 30 ($Q = -30$) at $j \sim 4.78 \times 10^{10}$ A m$^{-2}$ and then recovered to ~150 ($Q = -150$) at $j \sim 3.16 \times 10^{10}$ A m$^{-2}$. Therefore, we have shown that injecting and removing topological charges in the Fe$_3$Sn$_2$ microdevice can be realized by continuously decreasing and increasing the current density, respectively.

The residual counts (~20–40) for bubbles at small current density and for skyrmions at high current density are mainly contributed by magnetic objects near the two sides of the thin plate, whose transformations are not sensitive to current because of possible strong pinning effects near the sample edges. The non-coherent skyrmion-to-bubble transformation from the left side to the right side and the reverse transformation from the right side to the left side are understood by the current density inhomogeneity induced by thickness inhomogeneity of the plate. From the sharp transition current density range of ~3.88–4.16 × 10$^{10}$ A m$^{-2}$, we can approximately estimate that the thickness inhomogeneity of the Fe$_3$Sn$_2$ thin plate is ~ 7%, *i.e.*, (4.16–3.88) × 2/(4.16+3.88)×100%.

We further applied pulsed current with switched current density between two discrete values of ~3.79 and $4.43 \times 10^{10}$ A m$^{-2}$. The time interval during each switch of current density is ~90 s, which suggests that ~90 current pulses have been applied before each switch. We clearly observed a sudden increase in bubble count from ~40 to 120 along with a decrease in skyrmion count from ~100 ($Q = -100$) to 20 ($Q = -20$) upon the current density was switched from ~3.79 to $4.43 \times 10^{10}$ A m$^{-2}$ (**Figure 2**), which indicates the removal of topological charges. When switching the pulsed current density from ~4.43 back to $3.79 \times 10^{10}$ A m$^{-2}$, the skyrmion count gradually increased from ~20 ($Q = -20$) to ~100 ($Q = -100$) along with the bubble count decreasing from ~120 to ~40 after ~60 low current pulses were applied, which suggests the injection of topological charges (Figure 2). Importantly, the manipulation of topological charges is highly reversible and reproducible in the cycle of switching *j* (Figure 2a and Supplemental video 2), which suggests that injection and removal of topological charges in $Fe_3Sn_2$ thin plates can be flexibly and precisely realized by tuning current density and current pulses.

One prerequisite for the current-induced topological magnetic transformations between skyrmions and bubbles is the coexistence of skyrmion–bubble phases at a given field condition. Owing to different formation mechanisms of skyrmions and bubbles, skyrmions and bubbles are the only stable phases at normal field and large tilted field, respectively[26-32]. Therefore, for the $Fe_3Sn_2$ thin plate, the controlled current-induced skyrmion–bubble transformations can be realized in an intermediate tilted angular range (**Figure 3a**), such as from ~0.8° to 3.0° at $B \sim 500$ mT. Although skyrmions can be metastable phases even at zero field, bubbles merge together to form a long stripe domain in low field region at $B < \sim 400$ mT. Therefore, current-induced skyrmion–bubble transformations cannot be achieved in the low field region. However, we show that the manipulation of topological charges can be achieved by skyrmion–stripe transformations in the low field region (Figure 3 and Supplemental video 3). When continuously decreasing the current density *j* from ~4.88 to 3.61 × $10^{10}$ A m$^{-2}$ at $B \sim$ 375 mT with $\alpha \sim 3°$, skyrmions are generated by splitting striped domains and the skyrmion count increases from ~40 ($Q = -40$) to ~200 ($Q = -200$). Furthermore, a reversed process, *i.e.*, skyrmions annihilating to form striped domains and the skyrmion count decreasing from ~200 ($Q = -200$) to ~40 ($Q = -40$), occurs as the current density recovers back from ~3.61 to 4.88 × $10^{10}$ A m$^{-2}$. Therefore, current-induced injection and removal of topological charges are successfully achieved from skyrmion–stripe transformations in a reproducible and reversible way (Supplemental video 3). Note that although current-induced generation of skyrmions from helix or stripes has been previously reported, the reversible process for skyrmion annihilation has not been shown previously[18-25]. In the field range of ~400–460 mT, the manipulation of topological charges can be realized by coexisting skyrmion–bubble and skyrmion–stripe transformations (Supplemental Figure 4 and Supplemental video 4). As skyrmions with two helicities are in energetic equilibrium in the achiral uniaxial magnet $Fe_3Sn_2$, we do not observe an effect of the helicities on the current-generated efficiencies of skyrmions with different helicities (Supplemental Figure 5). Although we have realized various current-induced topological magnetic phase transitions in $Fe_3Sn_2$ crystal, the hardly coherent dynamical motion of skyrmions and bubbles may be attributed to bubble or skyrmion interactions among the densely arranged magnetic lattice and strong impurity pinning effects[28]. Topological trivial bubbles ($Q = 0$) easily transform to stripes ($Q = 0$) when the field is reduced without topological transformations. In contrast, nonvolatile topological nontrivial skyrmions ($Q = -1$) are observed even at zero field and zero current (Supplemental Figure 6).

Finally, we explore the microscopic physical mechanisms of the observed reversible topological magnetic transformations between skyrmions and bubbles using micromagnetic simulations[41]. The typical current effects include spin-transfer torque (STT) and Joule thermal heating[19, 25, 42, 43]. Here, we interpret the current-induced topological magnetic transformations in a controlled way by the two combined effects. Skyrmions and bubbles are both metastable phases in $Fe_3Sn_2$ separated by an energy barrier (Supplemental Figure 7)[44]. The skyrmion-to-bubble transformation at high current density and bubble-to-skyrmion transformation at low current density are explained by mainly the contribution of Joule thermal heating and STT effects, respectively. In $Fe_3Sn_2$, a previous study estimated that the temperature increases by ~180 K with a current of $j = 3.4 \times 10^{10}$ A m$^{-2}$ and a pulsed width of 100 ns based on COMOSOL Multiphysics simulations[28]. During 1 s duration between two 120-ns-width current pulses, the sample temperature will first rise when the pulse current is on and then return to room temperature in the 1 s interval before the next current pulse. Such heating and cooling process during each current pulse can induce a reduced saturated magnetization and magnetization recovery process. Using a micromagnetic simulation approach, we obtained the magnetic evolution induced by the reduced saturated magnetization and magnetization recovery process (**Figure 4b**, Supplemental Figure 8 and Supplemental video 5). The skyrmion gradually shrinks and finally transforms to a ferromagnetic state when the saturation magnetization $M_s$ at room temperature is reduced by 18.5% at a given tilted field condition. The reduced magnetization in $Fe_3Sn_2$ induced by an increased temperature is well experimentally identified and consistent with our simulations (Supplemental Figure 9). During the magnetization recovery by increasing $M_s$, we suggest that an energy barrier must be overcome to realize the transition from the ferromagnet to skyrmion as the in-plane magnetizations of tilted ferromagnet are against the closure domain wall magnetizations of the skyrmion. In contrast, the uniformed in-plane magnetizations of the tilted ferromagnet are collinear with the domain wall magnetizations of the bubble. Therefore, the tilted ferromagnetic state prefers to transform to the bubble state although its energy is higher than that of the skyrmion state. Such irreversible process is well numerically confirmed (Supplemental Figure 8 and video 5). During the field-driven magnetic evolution (Supplemental Figure 10), we observed the skyrmions shrink and finally transform to ferromagnet as field increases. When reducing field back to 500 mT, we obtained the bubble state. Such field-driven nucleation of bubbles transformed from the tilted ferromagnetic state (Supplemental Figure 10) coincides with proposed skyrmion-to-bubble transformation mechanism at high current density (Figure 4a). Therefore, Joule thermal heating and recovery process can well explain the skyrmion-to-bubble transformation at high current density.

However, at low current density, the maximum increase in temperature is not sufficiently large to achieve an intermediate ferromagnetic state, and consequently the skyrmion-to-bubble transformation is not favored. Of note, random helicity reversal happens at low current density (Supplemental Figure 3) and cannot be explained by pure thermal effect STT has been successfully applied to understand the current-induced helicity reversal of skyrmions in $Fe_3Sn_2$ crystal[28]. We suggest that the current-induced bubble-to-skyrmion transformations at low current density are mainly driven by a similar effect (Figure 4c and Supplemental video 6). STT is an external stimulation that facilitates the metastable bubble phase to overcome the energy barrier to form a skyrmion. In such process, impurity pinning effect also needs to be considered to pin the bubble when the bubble will otherwise move freely in the pinning-free system, and consequently no transformation occurs. When combing impurity pinning and STT, we show that the bubble moves slightly away from the pinning center, and then transforms to a high-energy state owing to the dragging torque from the pinning site. Finally the skyrmion phase are formed (Figure 4c) along with the switched topological magnetic charge $Q$ from 0 to $-1$ (Supplemental Figure 11). We also confirmed that the STT-driven bubble-to-skyrmion transformation can be realized after considering the experimental rising time (~3 ns) of pulsed current (Supplemental Figure 12). However, such bubble-to-skyrmion transformation depends on the current density; there exists a threshold current density of ~$1.9 \times 10^{11}$ A·m$^{-2}$ in zero-temperature simulation, below which the bubble-to-skyrmion transformation does not happen (Supplemental Figure 12d). In contrast, we reveal that the current density required for bubble-to-skyrmion transformation in experiments is in the range of ~$2.7$–$4.0 \times 10^{10}$ A·m$^{-2}$.

The discrepancy points out that thermal heating should also play a role in low-current density regions. For example, although STT by itself can explain well the bubble-to-skyrmion transformations, thermal fluctuation energy can also lower the transition energy barrier to realize the experimental bubble-to-skyrmion transformation at a small current density of the order of ~$10^{10}$ A m$^{-2}$. In addition, the thermal fluctuation energy plays an important role in stripe-to-skyrmion transformations at low field region to split the stripes[15]. We propose that the transformations from stripes to skyrmions at low field are attributed to the combined effects of Joule thermal heating and spin transfer torque. The transformation process from stripe to skyrmions can be considered as a two-step process (Supplemental Figure 13b): 1. The stripes first split up to form bubbles because of reduced saturated magnetizations induced by Joule thermal heating, which is numerically identified when saturated magnetization is reduced by 8%; 2. the high-energy metastable bubbles then transform to low-energy skyrmions driven by STT effect as shown in Figure 4b. The two-step stripe-to-skyrmion transformations are also experimentally identified (Supplemental Figure 13a).

**Conclusions**

In summary, using *in-situ* Lorentz TEM real-time magnetic imaging, we have shown current-controlled topological magnetic transformations between two distinct localized magnetic objects, *i.e.* skyrmions and bubbles, and generation and annihilation of skyrmions by skyrmion-stripe transformations. These topological magnetic transformations occur along with switched topological charges *Q* that determine the electromagnetic properties of magnetic domains. The room-temperature reproducible and reversible topological magnetic transformations achieved with current density of the order of ~$10^{10}$ A m$^{-2}$ enable future reliable and energy-efficient manipulations of electromagnetic properties. Thus, our findings should trigger future discoveries in topological spintronic device applications.

**Methods**

*Fabrication of Fe$_3$Sn$_2$ microdevices*: Thin Fe$_3$Sn$_2$ microdevices, comprised of a ~150-nm-thick region which is connected to two PtC$_x$ electrodes, were fabricated from the bulk using a standard lift-out method, with a focus ion beam and scanning electron microscopy dual beam system (Helios Nanolab 600i, FEI).

*TEM measurements*: High resolution Fresnel magnetic imaging was conducted using a TEM instrument (Talos F200X, FEI) operated at 200 kV and Lorentz mode. The variable out-of-plane field was realized by adjusting the object current. The current pulses with a pulsed width of 120 ns and frequency of 1 Hz were provided by a voltage source (AVR-E3-B-PN-AC22, Avtech Electrosystems Ltd.). The current-driven dynamics were measured at room temperature.

*Micromagnetic simulations*: The zero-temperature micromagnetic simulations were performed using a micromagnetic simulation program, Mumax3.[41] We consider in the Hamiltonian exchange interaction ($A$) energy, uniaxial magnetic anisotropy ($K_u$) energy, Zeeman energy, and dipolar-dipolar interaction energy.[41] Magnetic parameters are set based on the material $Fe_3Sn_2$ with $A$ = 8.25 pJ/m, $K_u$ = 54.5 kJ/m$^3$, and room-temperature saturation magnetization $M_s$ = 622.7 kA/m[26, 27]. The cell size was set at $2 \times 2 \times 3$ nm$^3$. The equilibrium spin configurations were obtained using the conjugate-gradient method. A Zhang-Li STT is considered for simulating current-driven dynamics.

**Supporting Information**

Supporting Information is available from the Wiley Online Library or from the author.

**Acknowledgements**

J. T. and Y. X. acknowledge the financial support of the Natural Science Foundation of China, Grant Nos. 11804343 and U1432138. H. D. acknowledges the financial support from the National Key R&D Program of China, Grant No. 2017YFA0303201; the Key Research Program of Frontier Sciences, CAS, Grant No. QYZDB-SSW-SLH009; the Major/Innovative Program of Development Foundation of Hefei Center for Physical Science and Technology, Grant No. 2016FXCX001; the Key Research Program of the Chinese Academy of Sciences, Grant No. KJZD-SW-M01; the Strategic Priority Research Program of Chinese Academy of Sciences, Grant No. XDB33030100; and the Equipment Development Project of Chinese Academy of Sciences, Grant No. YJKYYQ20180012. Y. X. acknowledges the financial support of the National Key R&D Program of China, Grant No. 2016YFA0300404; and the Collaborative Innovation Program of Hefei Science Center, CAS, Grant No. 2019HSC-CIP007.

**Author Contributions**

W.W. and J.T. contributed equally to this work. H.D. and J.T. supervised the project. J.T. conceived the idea and designed the experiments. Y.X., Y-H.W, and J.L. synthesized the $Fe_3Sn_2$ bulk crystals. W.W. fabricated the $Fe_3Sn_2$ microdevices. J.T., W.W., and Y-D.W. performed the TEM measurements. J.T. performed the simulations. J.T., H.D., and W.W. wrote the manuscript with input from all authors. All authors discussed the results and contributed to the manuscript.

**Data Availability Statement**

The data that support the findings of this study are available from the corresponding author on reasonable request.

**Conflict of Interest:**

The authors declare no competing financial interest.

**Figures:**

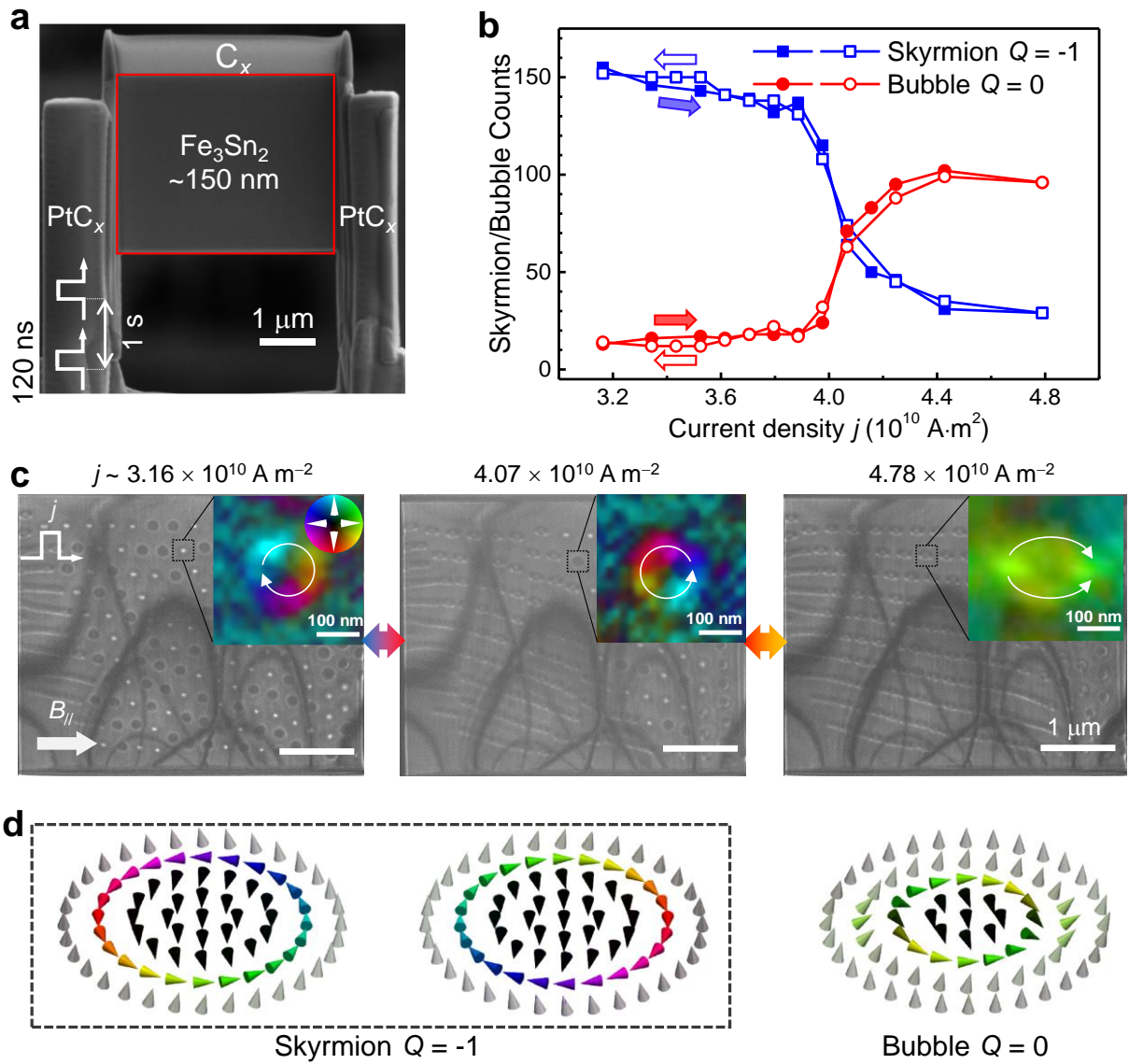

Figure 1. Skyrmion–bubble transformations controlled by continuously varied current density in an $Fe_3Sn_2$(001) thin plate. a) Scanning electron microscopy image of a microdevice comprised of a ~150-nm-thick $Fe_3Sn_2$ plate and two $PtC_x$ electrodes. An insulated carbon layer ($C_x$) was deposited on top of $Fe_3Sn_2$ as a protection layer. b) Counts of skyrmions with $Q = -1$ and bubbles with $Q = 0$ as a function of pulsed current density. The counts are obtained from the last snapshot before the next current density pulse. Filled and open symbols represent counts measured when increasing and decreasing the current density, respectively. c) Representative snapshots of defocused Fresnel contrasts at the pulsed current density $j \sim 3.16$, 4.07, and $4.78 \times 10^{10}$ A $m^{-2}$. Magnetic field ~500 mT with an in-plane field component $B_{//}$ ~

18 mT ($\alpha \sim 2.0°$) along the direction shown by the arrow. The insets of snapshots in (c) are in-plane magnetization mappings of a skyrmion with counter-clockwise rotation (white dotted Fresnel contrast), a skyrmion with clockwise rotation (black dotted Fresnel contrast), and a bubble (arc-shaped Fresnel contrast) obtained from differential phase contrast scanning transmission electron microscopy (DPC-STEM). The color in these insets represents the in-plane magnetization amplitude and orientation according to the colorwheel. d) Representative magnetic textures of skyrmions with two helicities and bubbles. Counter-clockwise and clockwise rotations of skyrmions with equivalent $Q = -1$ correspond to white and black dotted defocused Fresnel contrasts, respectively. The bubble typically shows an arc-shaped magnetic contrast. The defocused distance of Fresnel images is $-2$ mm.

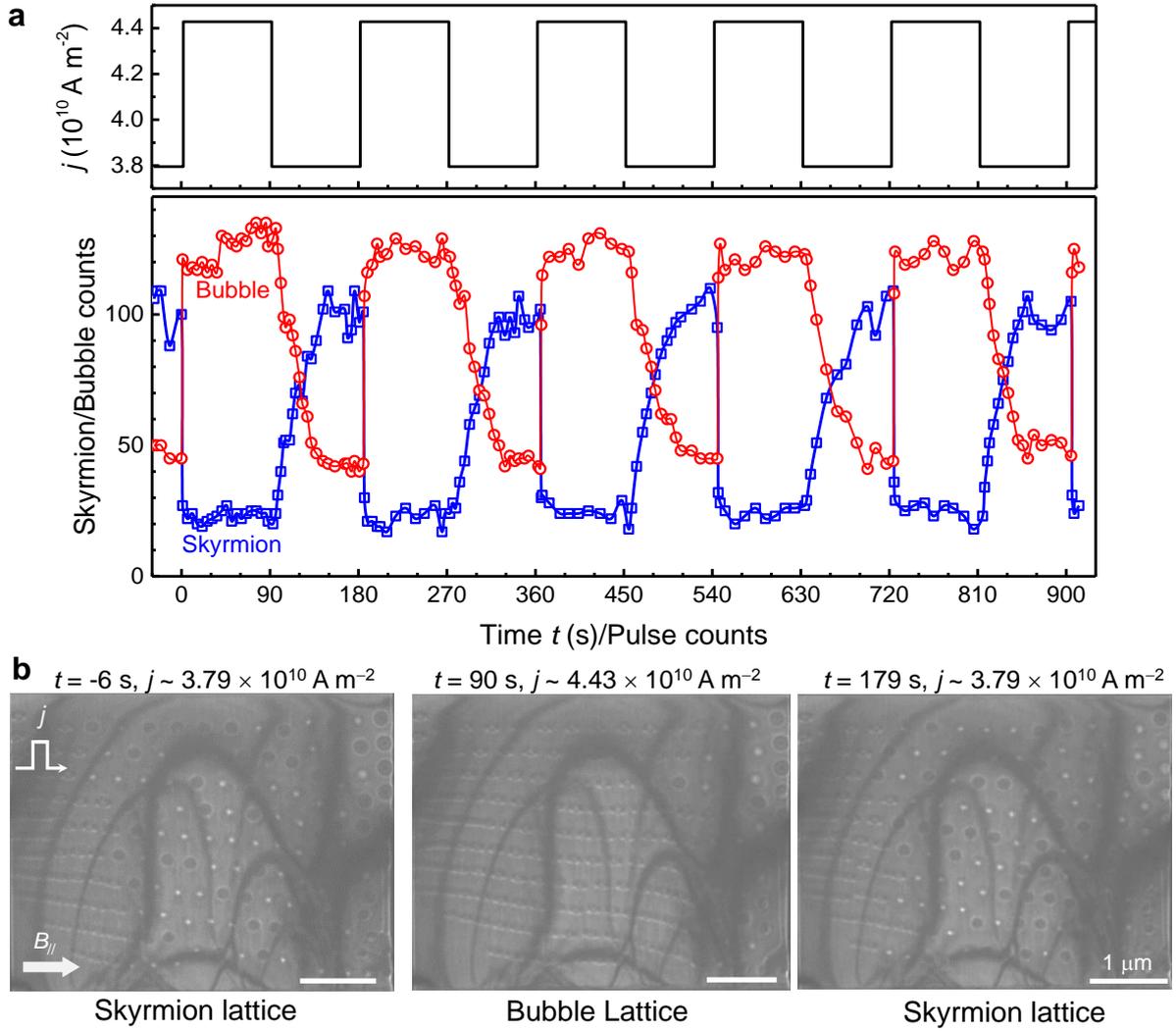

**Figure 2.** Current–controlled skyrmion–bubble transformations at two discrete current densities. a) Current density $j$ and corresponding counts of skyrmions and bubbles as a function of time $t$. We set the time interval between two pulsed current as 1 s. b) Snapshots of representative defocused Fresnel magnetic contrasts for the skyrmion-bubble transformations. Magnetic field ~500 mT with an in-plane field component $B_{//}$ ~ 18 mT ($\alpha$ ~ 2.0°) along the direction shown by the arrow in (b). The defocused distance of Fresnel images is –2 mm.

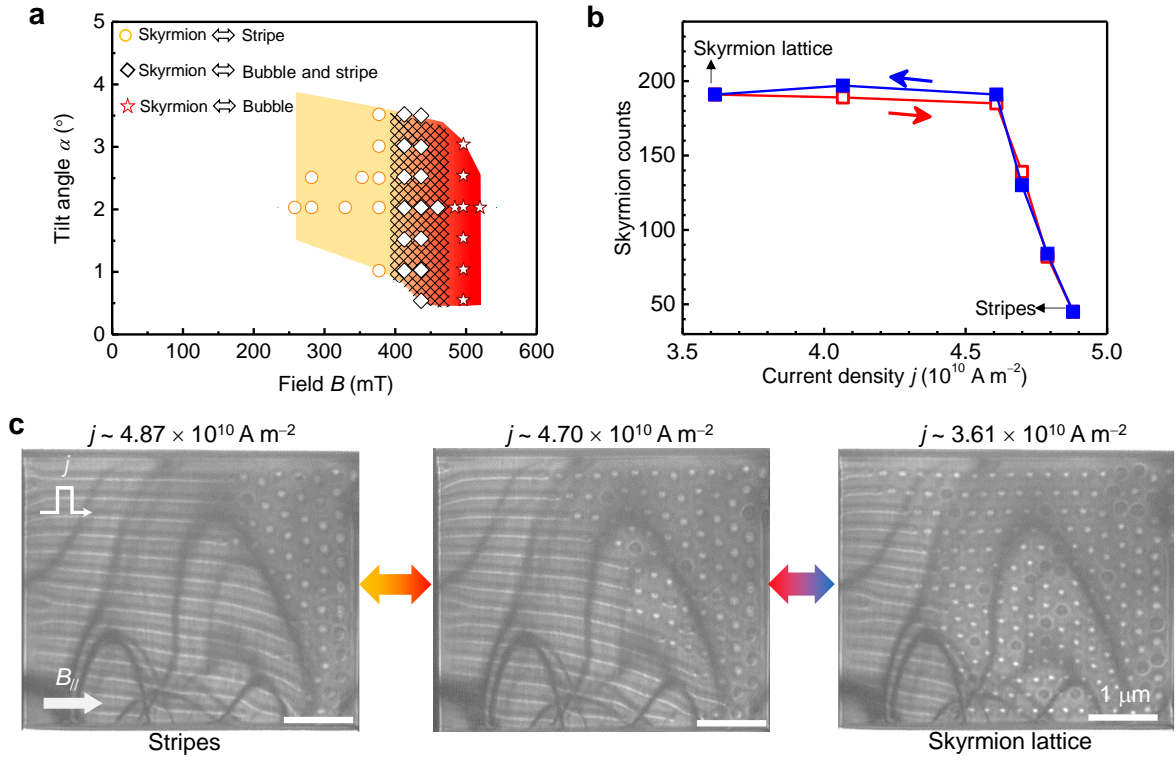

**Figure 3.** Current–controlled skyrmion-stripe transformations. a) Current-induced magnetic transition diagram as a function of field $B$ and tilted field angle $\alpha$ in the Fe$_3$Sn$_2$ thin plate. The red and yellow regions, which are guide for the eyes based on measured field conditions marked by symbols, represent the skyrmion–bubble (five-pointed stars) and skyrmion–stripe (circle dots) transformations, respectively. The shadow region indicates the coexisting skyrmion-bubble and skyrmion-stripe transformations (square dots). b) Skyrmion count as a function of current density during the skyrmion–stripe transformation. c) Snapshots of the representative skyrmion–stripe transformation process at current density $j \sim 4.87$, $4.70$, and $3.61 \times 10^{10}$ A m$^{-2}$. Magnetic field condition for (b) and (c) as marked is $B \sim 375$ mT and $\alpha \sim 3°$, with an in-plane field component $B_{//} \sim 20$ mT along the direction shown by the arrow in (c). The defocused distance of Fresnel images is $-2$ mm.

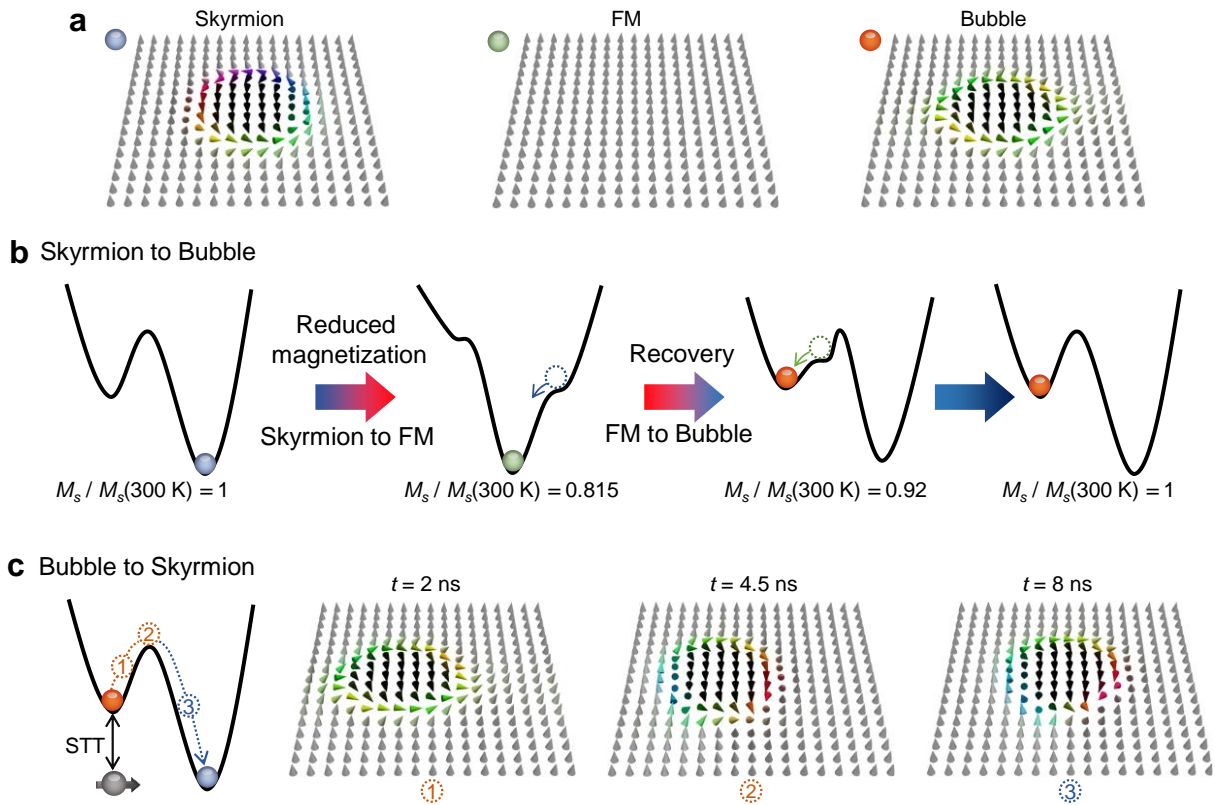

**Figure 4.** Simulated skyrmion–bubble transformations. Magnetic field $B$ = 400 mT and tilted field angle $\alpha$ = 1.2° a) Schematic configurations of the skyrmion, ferromagnet (FM), and bubble represented by blue, green, and orange balls, respectively. b) Schematic energetics during the topological skyrmion-to-bubble magnetic transformation induced by a reduced saturation magnetization and recovery process. c) Topological bubble-to-skyrmion transformation induced by a combined effect of STT and impurity. The numbers 1, 2, and 3 denote the intermediate configurations during the STT-driven transformations. Current density $j = 2.0 \times 10^{11}$ A m$^{-2}$.

**Current-Controlled Topological Magnetic Transformations in a Nanostructured Kagome Magnet**

*Wensen Wei[#], Jin Tang[#,*], Yaodong Wu, Yihao Wang, Jialiang Jiang, Junbo Li, Y. Soh, Yimin Xiong, Mingliang Tian, and Haifeng Du[*]*

Dr. W. Wei, Dr. J. Tang, Y. Wu, Dr. Y. Wang, J. Jiang, J. Li, Prof. Y. Xiong, Prof. M. Tian, Prof. H. Du
Anhui Province Key Laboratory of Condensed Matter Physics at Extreme Conditions, High Magnetic Field Laboratory, HFIPS, Anhui, Chinese Academy of Sciences, Hefei, 230031, China
[#]W.W. and J.T. contributed equally to this work.
[*]E-mail: tangjin@hmfl.ac.cn; duhf@hmfl.ac.cn

Y. Wu
Key Laboratory for Photoelectric Detection Science and Technology of Education Department of in Anhui Province, and School of Physics and Materials Engineering, Hefei Normal University, Hefei, 230601, China

Dr. Y. Soh
Paul Scherrer Institute, 5232, Villigen, Switzerland

Prof. M. Tian
School of Physics and Materials Science, Anhui University, Hefei, 230601, China

Prof. H. Du
Institutes of Physical Science and Information Technology, Anhui University, Hefei, 230601, China

**Supplemental Figures:**

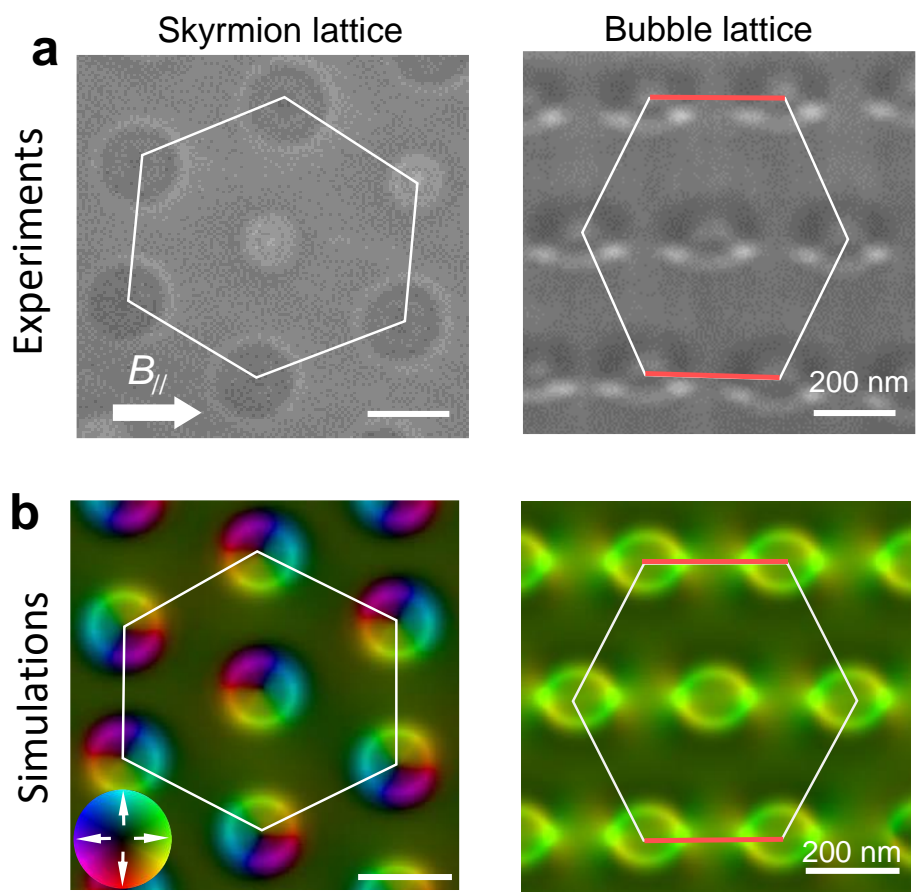

**Supplemental Figure 1**. a) A representative arrangement of skyrmions and bubbles obtained from Lorentz-TEM. b) Simulated in-plane magnetization mappings of skyrmion lattice and bubble lattice. $B$ = 420 mT, and tilted field angle $\alpha$ = 2°. The hexagonal unit cell of skyrmion (bubble lattice) is marked by a hexagon. Two sides of the hexagon in bubble lattice tend to line along with the in-plane field to minimize the dipole–dipole interaction. In contrast, the hexagon of skyrmion lattice has no preferable orientation.

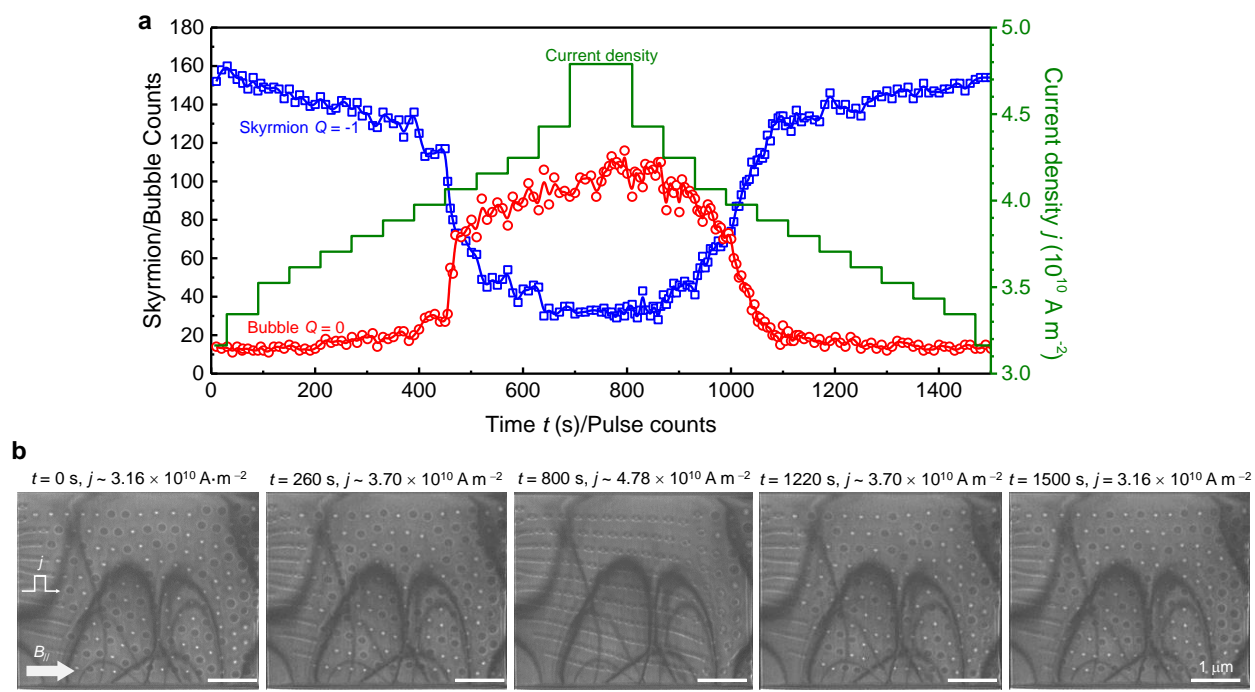

**Supplemental Figure 2.** Skyrmion–bubble transformations in the current density range of ~3.16–4.78× $10^{10}$ A m$^{-2}$. a) Skyrmion count, bubble count and current density as a function of time $t$. The frequency of pulsed current is 1 Hz. We applied ~60 pulses before each variation of current density $j$. b) Snapshots of defocused Fresnel magnetic configurations during the skyrmion-bubble transformation at time $t$ = 0, 260, 800, 1220, and 1500 s. Magnetic field $B$ ~ 500 mT with an in-plane field component $B_{//}$ ~ 18 mT ($\alpha$ ~ 2.0°). The figures are captured from the supplemental video 1.

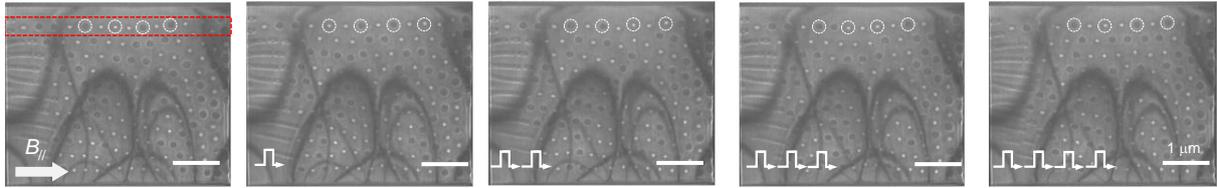

**Supplemental Figure 3.** Snapshots of current-induced random helicity reversals of skyrmions at $j \sim 3.16 \times 10^{10}$ A m$^{-2}$. Magnetic field $B \sim 500$ mT with an in-plane field component $B_{//} \sim 18$ mT ($\alpha \sim 2.0°$). The figures are captured from supplemental video 1.

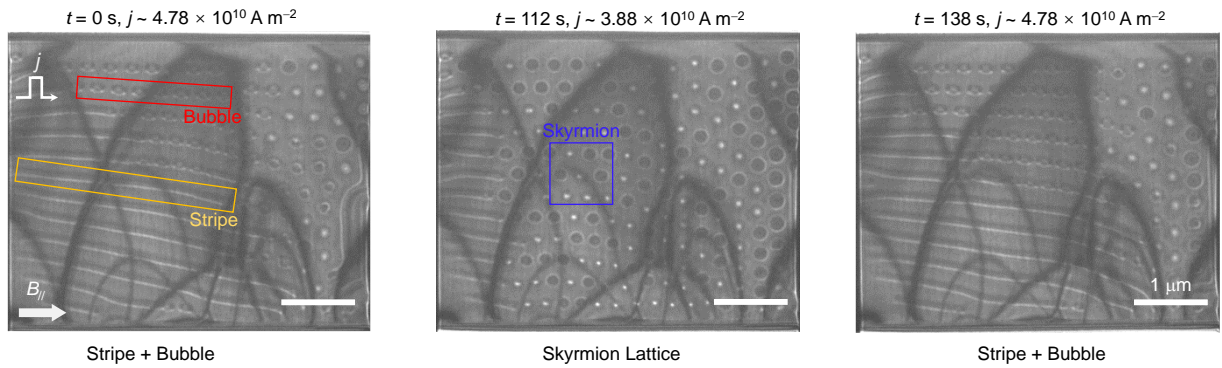

*t* = 0 s, *j* ~ 4.78 × $10^{10}$ A $m^{-2}$    *t* = 112 s, *j* ~ 3.88 × $10^{10}$ A $m^{-2}$    *t* = 138 s, *j* ~ 4.78 × $10^{10}$ A $m^{-2}$

Stripe + Bubble                Skyrmion Lattice                Stripe + Bubble

**Supplemental Figure 4.** Current-induced coexistence of skyrmion–bubble and skyrmion–stripe transformations at two discrete current density values of *j* ~ 4.78 × $10^{10}$ and 3.88 × $10^{10}$ A $m^{-2}$. Magnetic field *B* ~ 460 mT with an in-plane field component $B_{//}$ ~ 16 mT ($\alpha$ ~ 3.0°) along the direction depicted by the arrow. The figures are captured from Supplemental Video 4.

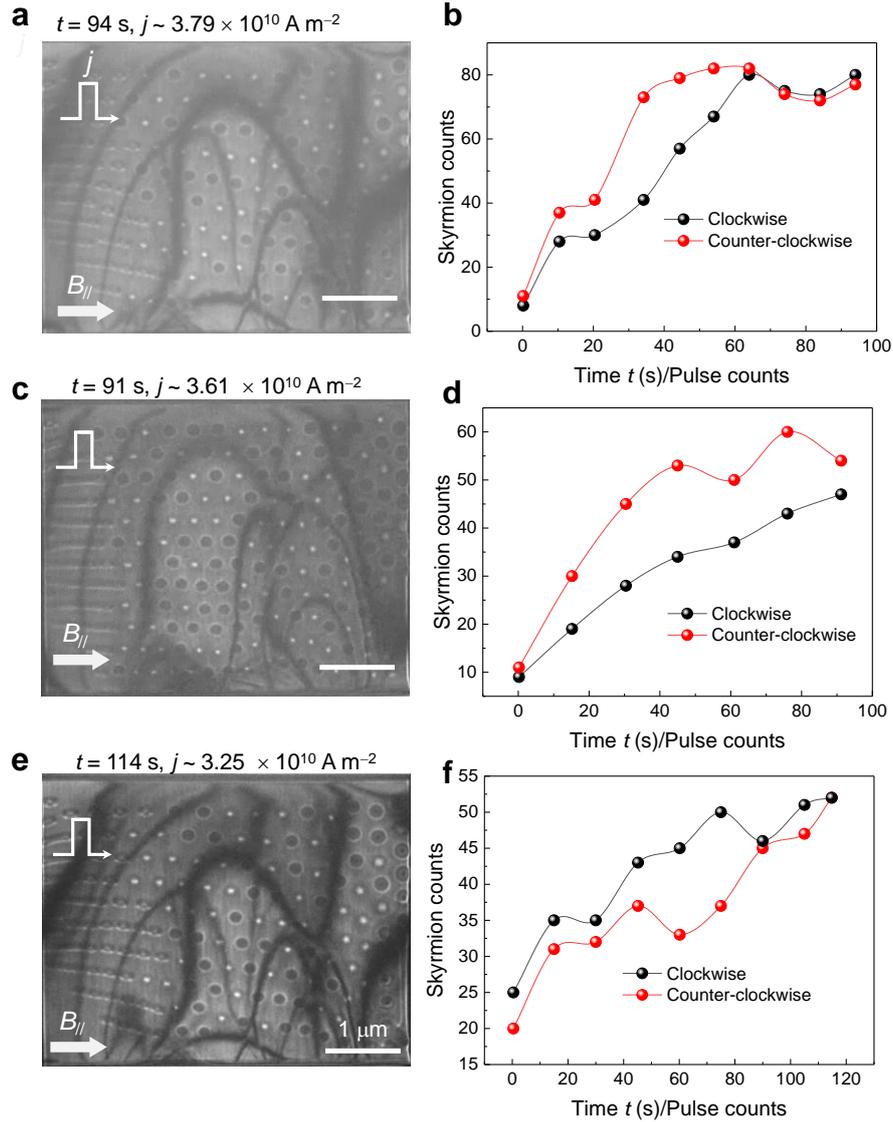

**Supplemental Figure 5**. a, b) Current-generated skyrmions after applying 94 pulsed currents at $j \approx 3.79 \times 10^{10}$ A m$^{-2}$ from an initial bubble state. The clockwise and counter-clockwise rotated skyrmion counts as a function of pulse count. $B \approx 500$ mT and $\alpha \approx 2°$. c, d) Current-generated skyrmions after applying 91 pulsed currents at $j \approx 3.61 \times 10^{10}$ A m$^{-2}$ from an initial bubble state. The clockwise and counter-clockwise rotated skyrmion counts as a function of pulse count. $B \approx 170$ mT and $\alpha \approx 2.5°$. e, f) Current-generated skyrmions after applying 114 pulsed currents at $j \approx 3.25 \times 10^{10}$ A m$^{-2}$ from an initial bubble state. The clockwise and counter-clockwise rotated skyrmion counts as a function of pulse count. $B \approx 500$ mT and $\alpha \approx 2°$. The helicities of current-generated skyrmions do not show a general tendency.

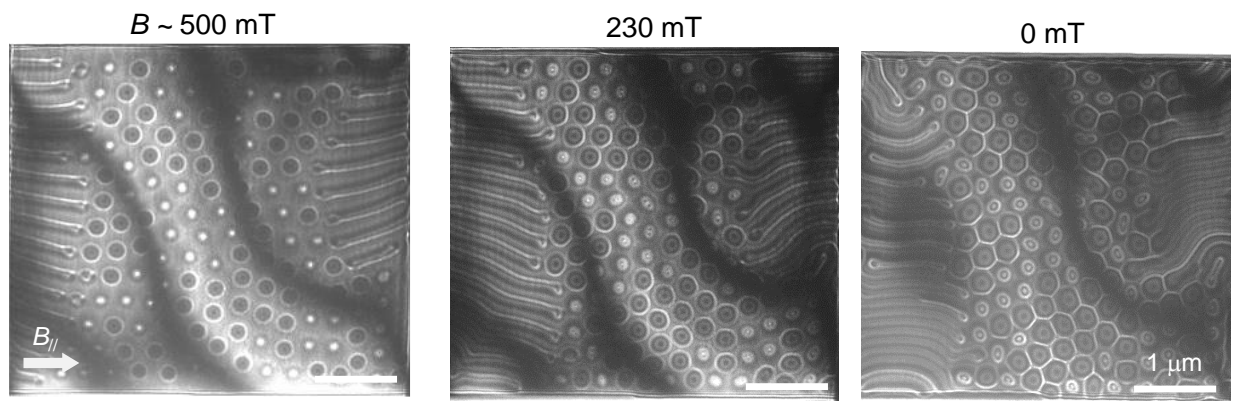

**Supplemental Figure 6**. Nonvolatility of skyrmions at zero field and zero current. Zero-field skyrmions obtained by reducing field from $B \approx 500$ mT. Tilted field angle $\alpha \approx 3°$.

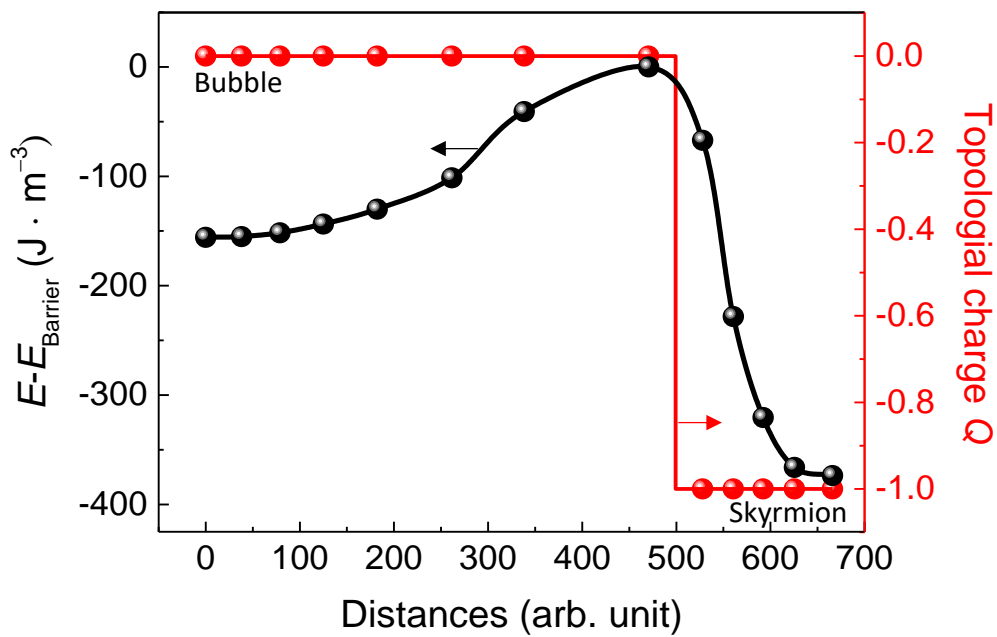

**Supplemental Figure 7.** Energy barriers for bubble-to-skyrmion transformations obtained from numerical geodesic nudged elastic band simulations. The barrier is taken at the position where the energy reaches maximum $E_{\text{Barrier}}$. The corresponding topological charge $Q$ is also shown in the figure. Magnetic field $B$ = 400 mT and tilted angle $\alpha$ = 1.5°.

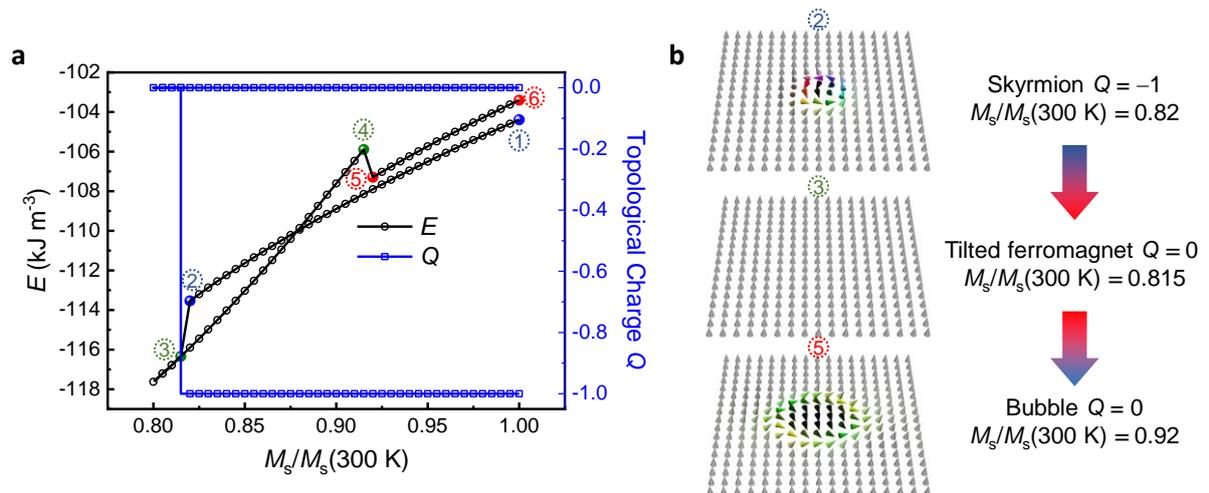

**Supplemental Figure 8.** Simulated transformation process from the skyrmion to the bubble achieved by reduced saturation magnetization and magnetic recovery. a) Total energy density $E$ and topological charge $Q$ as a function of saturation magnetization $M_s$. The transition process follows the sequence numbered from 1 to 6, whereby 1 to 2 refer to the skyrmion, 3 to 4 refer to the tilted ferromagnet, and 5 to 6 refer to the bubble. b) Representative magnetic textures during the skyrmion-to-bubble transformation. Magnetic field $B$ = 400 mT, and tilted field angle $\alpha$ = 1.2°.

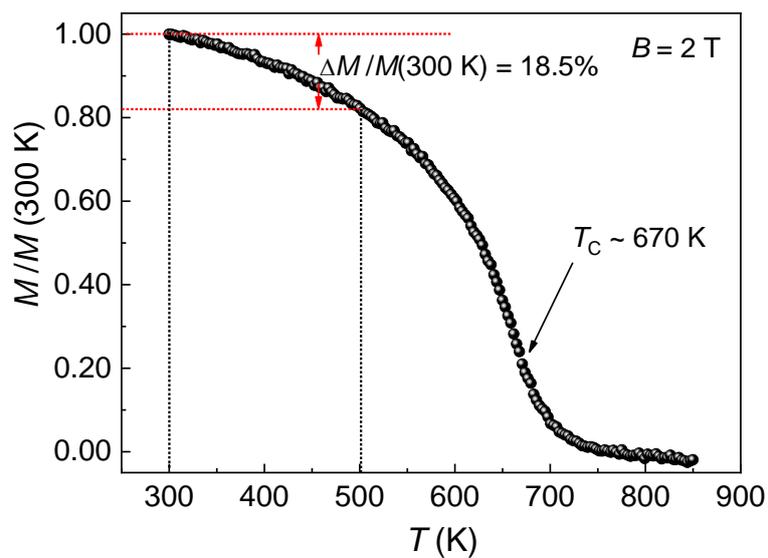

**Supplemental Figure 9**. Temperature dependence of magnetization at $B = 2$ T applied perpendicular to the $c$ axis. We identified a reduced magnetization of 18.5% at ~500 K from 300 K.

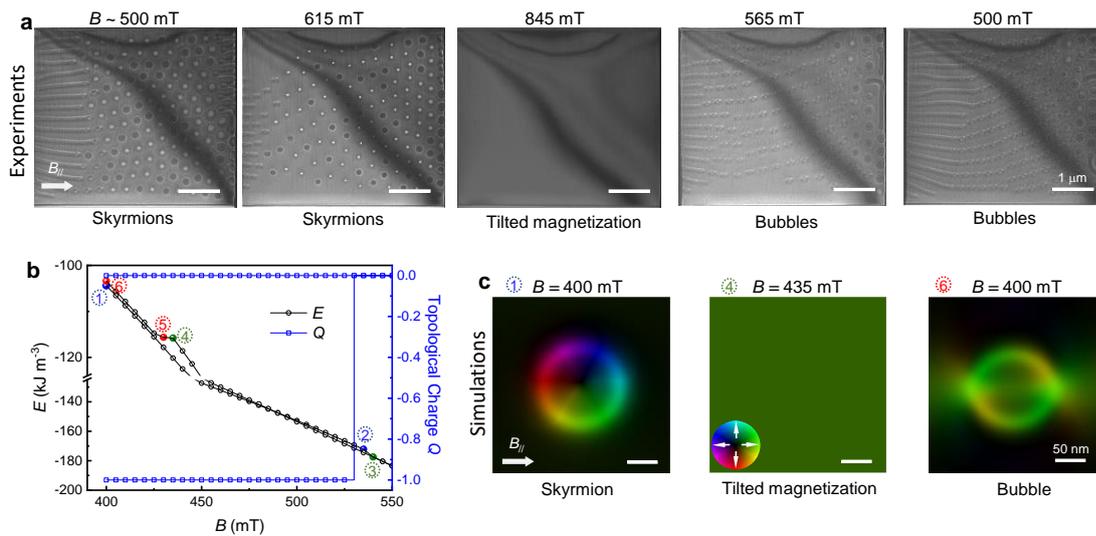

**Supplemental Figure 10**. a) Experimental magnetic field driven skyrmion-to-bubble transformation with an intermediate tilted ferromagnetic state. b) Simulated field-dependence of energy profile. The transition process follows the sequence numbered from 1 to 6, whereby 1 to 2 refer to the skyrmion, 3 to 4 refer to the tilted ferromagnet, and 5 to 6 refer to the bubble. c) Simulated in-plane magnetization mappings during the field-driven skyrmion-to-bubble transformation. Since the tilted in-plane magnetization supports the domain wall magnetization of the bubble, the bubble is the preferred state despite its higher energy than that of the skyrmion by transformation from tilted ferromagnet. The color represents the in-plane magnetization orientation and amplitude according to the colorwheel. Tilted angle $\alpha = 1.2°$.

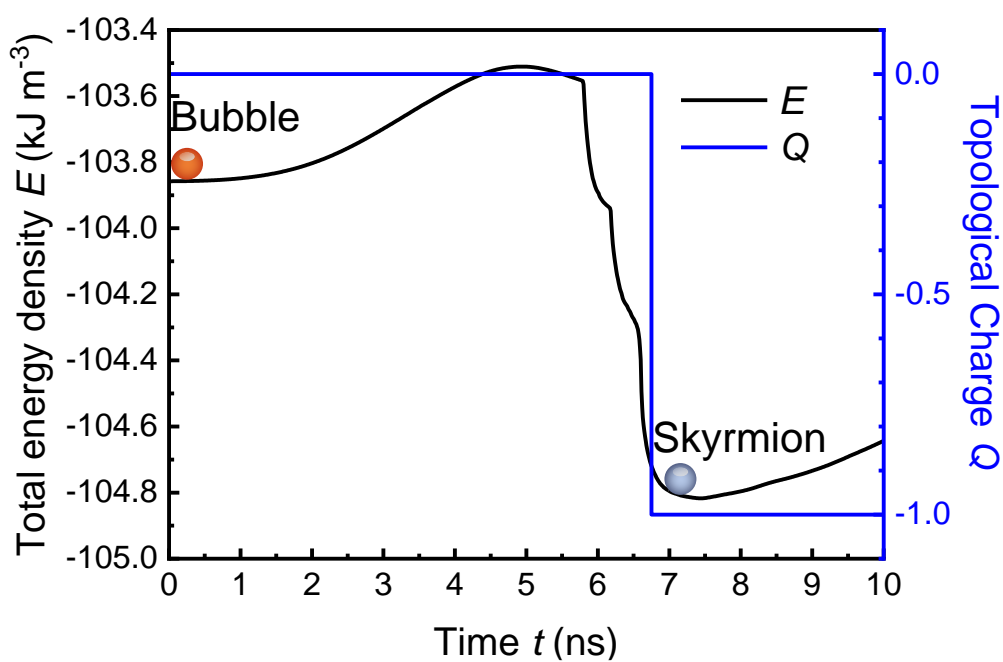

**Supplemental Figure 11**. Simulated total energy density and topological charge $Q$ as a function of time $t$ for the bubble-to-skyrmion transformation induced by spin transfer torque. The core of the magnetic object is initially pinned by a pinning site with a higher anisotropy $K_{\text{pinning}} = 15K_u$. $j = 2.0 \times 10^{10}$ A m$^{-2}$, magnetic field $B = 400$ mT, and tilted field angle $\alpha = 1.2°$.

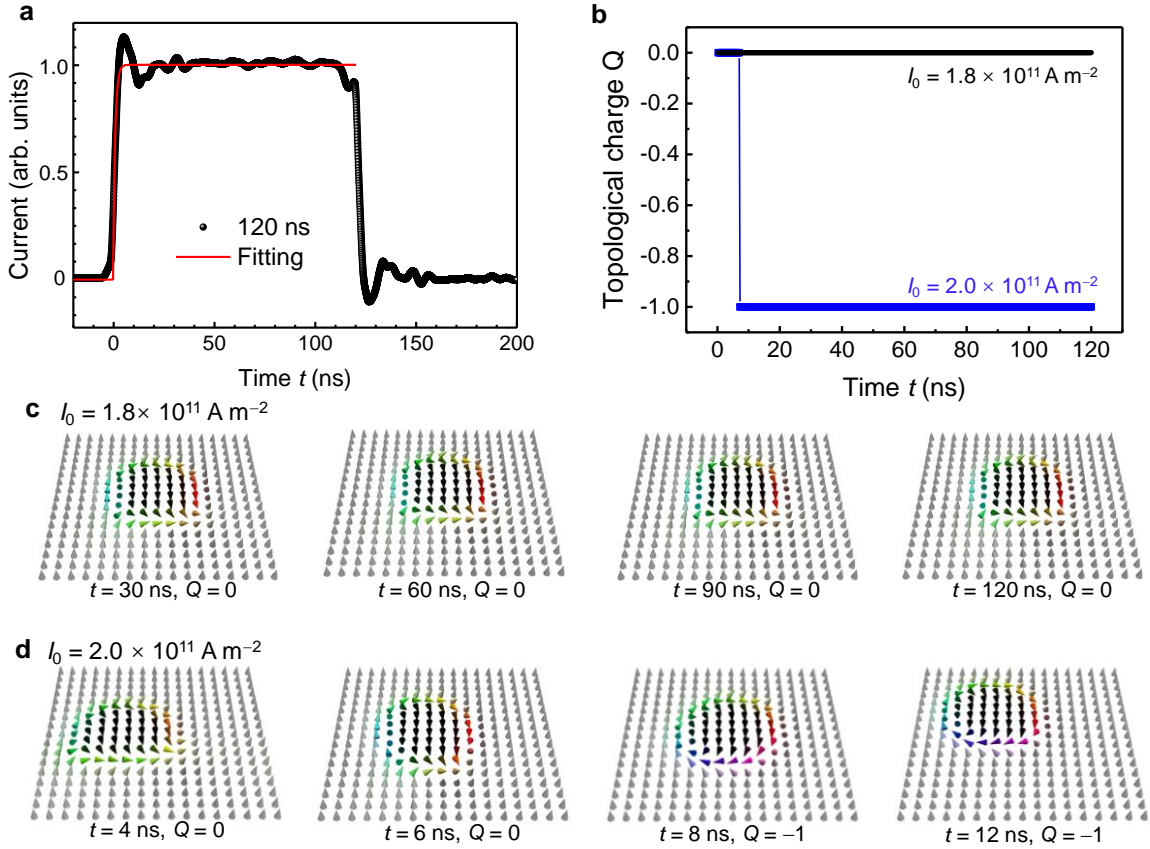

**Supplemental Figure 12**. Current-driven bubble-to-skyrmion transformation after considering a rising time of pulsed current. a) Measured 120-ns pulsed current profile. The black and dotted results are the experimental and fitting results with a function of $I_0[1-\exp(-2\pi f t)]$ with $f$ = 0.164 GHz, respectively. $I_0$ is the set pulsed current amplitude. b) Simulated time dependence of topological charge $Q$ after applying the fitted current density on an initial bubble state at $I_0 = 2.0 \times 10^{11}$ A m$^{-2}$ and $1.8 \times 10^{11}$ A m$^{-2}$. c) Representative simulated magnetic textures at $I_0 = 1.8 \times 10^{11}$ A m$^{-2}$. The bubble cannot transform to the skyrmion at low current density. d) Representative magnetic textures during the bubble-to-skyrmion transformation at $I_0 = 2.0 \times 10^{11}$ A m$^{-2}$. Once the skyrmion is stabilized, the skyrmion keeps stable and no topological transformation occurs afterward. The core of the magnetic object is initially pinned by a pinning site with a higher anisotropy $K_{\text{pinning}} = 15K_u$. Magnetic field $B$ = 400 mT and tilted field angle $\alpha$ = 1.2°.

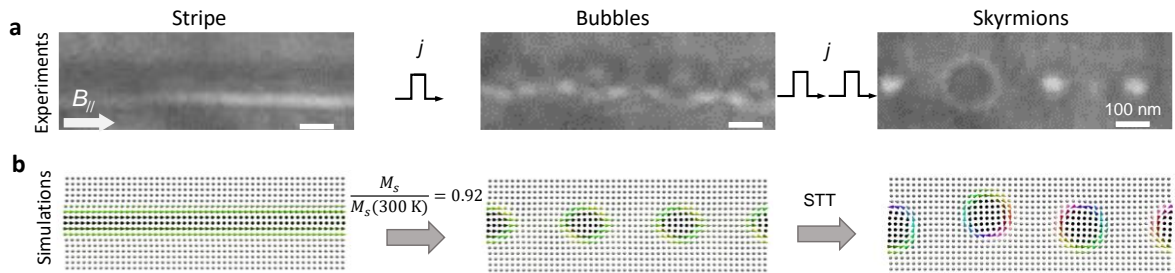

**Supplemental Figure 13**. a) A representative two-step process for stripe-to-skyrmion transformation. The stripe first transforms to bubbles after the first pulse and then transforms to skyrmions after the second pulse. $B \sim 460$ mT, and tilted field angle $\alpha \sim 2°$. b) Corresponding simulated stripe-to-skyrmion transformation. The first step from stripe-to-bubble is attributed to the reduced saturated magnetization ($M_s$) induced by Joule thermal heating. The second step from bubbles to skyrmions is attributed to the spin transfer torque (STT). $B = 350$ mT, and tilted field angle $\alpha = 1.5°$.

**Supplemental Video Captions:**

**Supplemental Video 1**. Current density dependence of magnetic dynamics in the $Fe_3Sn_2$ thin plate at $B \sim 500$ mT with an in-plane field component $B_{//} \sim 18$ mT ($\alpha \sim 2.0°$). The current density $j$ increases gradually from ~3.16 to $4.78 \times 10^{10}$ A m$^{-2}$ and then recovers to $3.16 \times 10^{10}$ A m$^{-2}$. We applied ~60 pulses before each variation of current density $j$. The time $t$ in the video is sped up by fifteen times.

**Supplemental Video 2**. Controlled current-induced skyrmion−bubble transformations at $B \sim 500$ mT with an in-plane field component $B_{//} \sim 18$ mT ($\alpha \sim 2.0°$). The current density $j$ is switched between two discrete values of ~3.79 and $4.43 \times 10^{10}$ A m$^{-2}$. We applied ~90 pulses before each variation of current density $j$. The time $t$ in the video is sped up by ten times.

**Supplemental Video 3**. Controlled current-induced skyrmion−stripe transformations at $B \sim 375$ mT with an in-plane field component $B_{//} \sim 20$ mT ($\alpha \sim 3.0°$). The current density $j$ decreases gradually from ~4.88 to $3.61 \times 10^{10}$ A m$^{-2}$ and then recovers to $4.88 \times 10^{10}$ A m$^{-2}$. We applied ~60 pulses before each variation of current density $j$. The time $t$ in the video is sped up by fifteen times.

**Supplemental Video 4**. Controlled current-induced coexisting skyrmion−stripe and skyrmion−bubble transformations at $B \sim 460$ mT with an in-plane field component $B_{//} \sim 16$ mT ($\alpha \sim 2.0°$). The current density $j$ is switched between two discrete values of ~3.78 and $4.88 \times 10^{10}$ A m$^{-2}$. We kept the initial state with current $j \sim 4.88 \times 10^{10}$ A m$^{-2}$, then, applied ~120 pulses for $j \sim 3.78 \times 10^{10}$ A m$^{-2}$ before switching the current density to $j \sim 4.88 \times 10^{10}$ A m$^{-2}$. The time $t$ in the video is sped up by five times.

**Supplemental Video 5**. Simulated skyrmion-to-bubble transformation induced by demagnetization and magnetization recovery effects. $B = 400$ mT with an in-plane field component $B_{//} = 10$ mT.

**Supplemental Video 6**. Simulated current-driven bubble-to-skyrmion transformations at $B = 400$ mT with an in-plane field component $B_{//} = 10$ mT and current density $j = 2.0 \times 10^{11}$ A m$^{-2}$. The core of the magnetic object is initially pinned due to a pinning site with a higher anisotropy $K_{\text{pinning}} = 15 K_u$.